\documentclass[twocolumn,prl,amsmath,amssymb,showpacs]{revtex4-2}
\usepackage{epsf}
\usepackage[utf8]{inputenc}
\usepackage{amsmath}
\usepackage{amsfonts}
\usepackage{amssymb}
\usepackage{makeidx}
\usepackage{color}
\usepackage{gensymb}
\usepackage{graphicx}

\begin{document}
\title {\mbox{{Structural, magnetic and transport properties of Co$_2$CrAl epitaxial thin films}}}
\author{Guru Dutt Gupt}
\affiliation{Department of Physics, Indian Institute of Technology Delhi, Hauz Khas, New Delhi-110016, India}
\author{Rajendra S. Dhaka}
\email{rsdhaka@physics.iitd.ac.in}
\affiliation{Department of Physics, Indian Institute of Technology Delhi, Hauz Khas, New Delhi-110016, India}
	
	\date{\today}       
		
	\begin{abstract}
We report the physical properties of Co$_2$CrAl Heusler alloy epitaxial thin films grown on single crystalline MgO(001) substrate using pulsed laser deposition technique. The x-ray diffraction pattern in $\theta$-2$\theta$ mode showed the film growth in single phase B2-type ordered cubic structure with the presence of (002) and (004) peaks, and the film oriented along the MgO(001) direction. The $\phi$~scan along the (220) plane confirms the four-fold symmetry and the epitaxial growth relation found to be Co$_2$CrAl(001)[100]$\vert$$\vert$MgO(001)[110]. The thickness of about 12~nm is extracted through the analysis of x-ray reflectivity data. The isothermal magnetization (M--H) curves confirm the ferromagnetic (FM) nature of the thin film having significant hysteresis at 5 and 300~K. From the in-plane M--H curves, the saturation magnetization values are determined to be 2.1~$\mu$$_{\rm B}$/f.u.~at 5~K and 1.6~$\mu$$_{\rm B}$/f.u. at 300~K, which suggests the soft FM behavior in the film having the coercive field $\approx$ 522~Oe at 5~K. The thermo-magnetization measurements at 500~Oe magnetic field show the bifurcation between field-cooled and zero-field-cooled curves below about 100~K. The normalized field-cooled magnetization curve follows the T$^2$ dependency, and the analysis reveal the Curie temperature around 335$\pm$11~K. Moreover, the low-temperature resistivity indicates semiconducting behavior with the temperature, and we find a negative temperature coefficient of resistivity (5.2 $\times$ 10$^{-4}$ /K). 
		
	\end{abstract}
	
	\maketitle 
	
	\section{\noindent ~INTRODUCTION}
	
	The full Heusler alloys having chemical formula as X$_2$YZ, where X and Y are the transition elements and Z is the sp-block element \cite{Felser_sips_16}, have attracted much attention to the scientific community for their wide range of potential applications. For example, in spintronic devices~\cite{Zutic_rmp_04, Hirohata_jpdap_14}, magnetic tunnel junctions (MTJ)~\cite{Wen_pra_14, Conca_jpdap_07}, magnetic refrigerators \cite{Nehla_prb_19, Nehla_jap_19}, and spin transistors~\cite{Caballero_jvst_98} because various unique properties of half metallicity, compensated ferrimagnetism and magnetocaloric effect~\cite{Groot_prl_83, Srivastava_aip_20, Nehla_prb_19, Nehla_jap_19}. Also, these compounds exhibit a high magnetic moment per unit cell and high Curie temperature ($\ge$ 300~K), which are essential for the stable device operation  \cite{Felser_sips_16, Nehla_prb_19, Guillemard_jap_20}. The half-metallic property of Heusler alloys leads to 100\% spin polarization for which conduction is metallic only for one spin band and insulating for the other spin band with a gap at the Fermi level (E$_{\rm F}$) \cite{Felser_sips_16, Husmann_prb_06, Groot_prl_83}. Remarkably, the half-metallic ferromagnetism (HMF) observed in Co$_2$-based alloys where the Co$_2$CrAl considered most important due to its tunable magnetic moment, high spin-polarization and Curie temperature just about 300~K \cite{Kudryavtesev_prb_08, Pan_pra_17, Galanakis_jpcm_02}. These alloys are crystallized in four interpenetrating face-centered cubic sub-lattices belong to the higher symmetry L2$_1$-type structure and in some cases lower symmetry structures like B2 (Pm$\bar{3}$m), A2 (Im$\bar{3}$m) and DO$_3$ (Fm$\bar{3}$m) \cite{Felser_sips_16}. Based on the band structure calculation, the value of the magnetic moment of  Co$_2$CrAl was theoretically reported to be 3~$\mu$$_{\rm B}$/f.u., which found to be consistent with the Slater-Pauling rule \cite{Felser_sips_16, Galanakis_prb_02}. However, the experimentally observed value is always much less due to several factors like disorders and/or complex magnetic interactions \cite{Nehla_prb_19, Kudryavtesev_prb_08}. 
	
	Note that most of the studies on Heusler alloys are on bulk alloys; however, it is important to investigate their physical properties in the form of thin films on different substrates for device fabrication \cite{Wen_pra_14, Hirohata_jpdap_14, Conca_jpdap_07, Guillemard_jap_20}. A few Heusler alloy thin films were grown on single-crystalline substrates such as Si, MgO, STO, MgAl$_2$O$_4$, and GaAs using molecular beam epitaxy \cite{Ambrose_jap_2000, Holmes_apl_02, Yamada_prm_18, Yamada_prb_19} or sputtering \cite{Geierbach_tsf_02, Kelekar_jap_04, Alfonsov_prb_15, Nakajima_jap_05}. In this context, few groups reported thin films of Co$_2$CrAl and found the Curie temperature around $\approx$330~K on single crystalline substrates, and also deposited on the glass substrate using flash evaporation method \cite{Kudryavtesev_prb_08, Kelekar_jap_04}. The thin films of Co$_2$CrAl grown on glass, NaCl substrate possesses the polycrystalline nature and grown on single crystalline MgO substrate having the epitaxial growth as well as the B2-type ordered structure \cite{Kudryavtesev_prb_08, Kelekar_jap_04}. However, pulsed laser deposition (PLD) is considered a simple technique with unique advantage of stoichiometric growth of multi-elemental films having wide range of vapor pressures of elements in the alloys \cite{Anupam_jpdap_10}. The PLD has been used to deposit the Heusler thin films on different semiconducting substrates to achieve the best stoichiometry ratio and good quality polycrystalline as well as oriented epitaxial thin films \cite{Wang_prb_05, Valerio_ass_05, Kushwaha_apl_17, PatraJALCOM19}.

	Moreover, the resistivity behavior of Co$_2$CrAl alloy has been investigated by several other groups with different atomic disordered structure~\cite{Husmann_prb_06, Kelekar_jap_04, Kudryavtsev_epjb_12} where it shows a typical metallic nature for the perfectly ordered L2$_1$ structure \cite{Zhang_jmmm_04} and a negative temperature coefficient of resistivity (TCR) for the B2-type ordered sample \cite{Husmann_prb_06, Kelekar_jap_04}. It is worthy to note that as a general rule, the half-metallic ferromagnets show a high value of residual resistivity ($\rho$$\rm _0$); for example, 1.5~$\mu$$\ohm$-m for bulk Co$_2$CrAl and 1.3~$\mu$$\ohm$-m for the Co$_2$CrGa \cite{Kourov_cap_15}. Interestingly, the value of negative TCR is reported for Co$_2$CrAl (-0.134~$\mu$$\ohm$cm/degree) \cite{Kudryavtsev_epjb_12} and CoFeMnSi (--7 $\times$ 10$^{–10}$~$\ohm$m/K) \cite{Kushwaha_apl_17}. The origin of the negative TCR can be either localization of the charge carrier or the presence of atomic disorder in the sample \cite{Kudryavtesev_prb_08, Kelekar_jap_04}. 
		
	Therefore, in order to establish the correlations between structural, magnetic, and transport properties, we successfully deposited epitaxial thin film of the Co$_2$CrAl Heusler alloy on a single-crystalline MgO(001) substrate using the pulsed laser deposition technique \cite{Shukla_tsf_20}. The XRD pattern and $\phi$-scan confirm the good quality film and epitaxial growth. The thickness extracted using XRR analysis is found to be around 12~nm. The thermo-magnetization (M--T) measurement at 500~Oe magnetic field shows the Curie temperature around 335$\pm$11~K, which is consistent with the fitting of field-cooled curve using empirical power law. The isothermal magnetization (M--H) curves recorded as a function of the magnetic field at 5~K as well as 300~K confirm ferromagnetic nature of the film sample. The saturation magnetization value is found to be 2.1~$\mu$$_{\rm B}$/f.u. at 5~K and 1.64~$\mu$$_{\rm B}$/f.u. at 300~K. Moreover, the {\it dc} electrical resistivity increases at low temperatures along with a high residual resistivity ($\approx$311~$\mu$$\ohm$-cm) value of the Co$_2$CrAl thin film. Thus, the observed semiconducting nature of the thin-film is useful in spin injector devices at room temperature. 
	
	\section{\noindent ~Experimental}
	
We use arc melting system (from Centorr vacuum industries, USA) to prepare the target of Co$_2$CrAl alloy by taking the  amount of pure elements of Co (Alfa Aesar, 99.9\%), Cr (Sigma, 99.995\%) and Al (Alfa Aesar, 99.999\%) in stoichiometric ratio, more details can be found in \cite{Nehla_prb_19,Nehla_jap_19}. The thin films of Co$_2$CrAl are deposited on a single-crystalline MgO(001) substrate using pulsed laser deposition with the base pressure in the range of good 10$^{-9}$ mbar in the deposition chamber \cite{Shukla_tsf_20}. Prior to the film deposition, the substrate was outgassed {\it in-situ} for 1 hr at 800$^{\rm o}$C using a resistive heater to improve the quality of the surface. A KrF excimer laser (from Coherent Inc.) of wavelength 248~nm was used for deposition having high laser fluence with an energy density of 2.5 J/cm$^2$ and fix repetition rate of 10~Hz. The focused laser beam incident at an angle of 45$^{\rm o}$ on the rotating and toggling target to avoid any plume disorientation and to make sure that the surface of the target remains homogeneous during the ablation \cite{Shukla_tsf_20}. The distance between the substrate to target was fixed at 5~cm and the pressure of 10$^{-7}$--10$^{-8}$ mbar was maintained during the deposition where the thin films of Co$_2$CrAl were grown at optimum temperature of 500$^{\rm o}$C to get the good quality film. After the deposition, the sample was annealed {\it in-situ} at 700$^{\rm o}$C for 30 minutes to enhance the crystallization and chemical ordering at thermodynamic equilibrium. 
	
	The structural phase and thickness of the film were analyzed at room temperature by x-ray diffraction (XRD) patterns in $\theta$-2$\theta$ mode and x-ray reflectivity (XRR) with Cu~K$\alpha$ radiation (1.5406~\AA) using the PANalytical x-ray diffractometer. The composition and the elemental mapping were done using the energy dispersive x-ray spectroscopy. The surface morphology and topography images were taken with field-emission secondary electron microscopy (FE-SEM) and atomic force microscopy (AFM). The temperature-dependent magnetization (M--T) and four-probe dc electrical resistivity ($\rho$$_{xx}$--T) measurements were performed using a physical property measurement system (PPMS) from Quantum Design, USA. We use a superconducting quantum interference device (SQUID) from Quantum Design, USA for collecting the in-plane isothermal magnetization (M--H) data. 
	
		\section{\noindent ~Results and Discussion}
	
	 The crystal structure of ordered L2$_1$ full-Heusler alloys Co$_2$CrAl consists of four interpenetrating {\it f.c.c.} sub-lattices having Co at 8$c$ (0.25, 0.25, 0.25), Cr at 4$b$ (0.5, 0.5, 0.5) and Al at 4$a$ (0, 0, 0) in the Wyckoff coordinate of the space group Fm$\bar{3}$m (no. 225)~\cite{Felser_sips_16, Graf_pssc_11}. 
	\begin{figure}[ht]
		\includegraphics[width=3.4in]{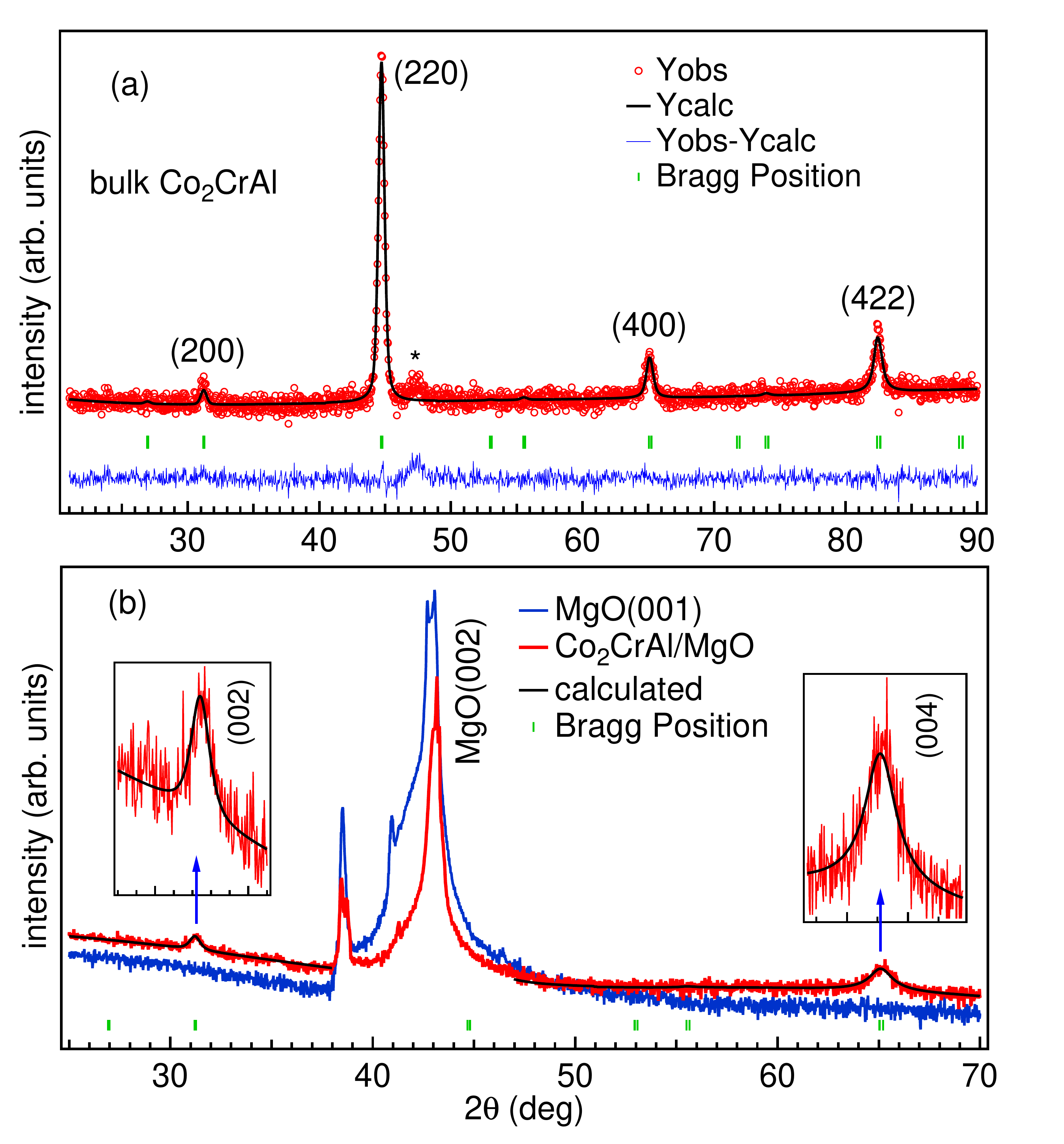}
		\caption{(a) The Rietveld refined x-ray diffraction (XRD) pattern of  Co$_2$CrAl bulk target, with a very small extra peak shown by black asterisk tag. (b) The Gonio XRD patterns of Co$_2$CrAl thin film and single-crystalline MgO(001) substrate in $\theta$-2$\theta$ mode (out of plane) where the insets highlight the Rietveld refinement of the film XRD peaks only. } 
\label{fig1}
\end{figure}
	Generally, the presence of two superlattice reflections (111) and (002) towards the lower 2$\theta$ angles in the x-ray diffraction (XRD) pattern is sensitive to the nature of ordering in the sample \cite{Felser_sips_16,Graf_pssc_11}. For example, both of these reflections are present in case of fully ordered L2$_1$ structure; however, (111) reflection is normally found to be absent in case of the B2-type ordering. Here, in Fig.~\ref{fig1}(a), the Rietveld refinement of the XRD pattern of Co$_2$CrAl target indicates the B2-type ordering. Also, we consider the anti-site disorder between Cr and Al atoms during the Rietveld refinement, which further improved the fitting of XRD pattern where the refined lattice parameters are in agreement with ref.~\cite{Nehla_prb_19}. The crystal structure of Co$_2$CrAl film deposited on MgO(001) substrate is studied at room temperature by performing the XRD measurements in $\theta$--2$\theta$ mode (out of plane geometry) in the angle range of 25--70$^{\rm o}$ and compared with the bare substrate, as shown in Fig.~\ref{fig1}(b). The presence of (002) and (004) reflections at 2$\theta$=31.5$^{\rm o}$ and 65.8$^{\rm o}$, respectively, attributes to the formation of (001) oriented epitaxial film having the B2-type structure (random position of Cr and Al atoms), and no extra peaks confirm the phase purity of the thin film. The Rietveld refinement of the XRD pattern of thin film is performed using the pseudo-Voigt function, excluding the 2$\theta$=38$^{\rm o}$--47$^{\rm o}$ from the fitting in the Fullprof software due to the presence of strong substrate related peaks in this range. The lattice constant of the Co$_2$CrAl thin film is found to be 5.731~\AA, which is the best match with the bulk lattice constant (5.730~\AA) as determined from Fig.~\ref{fig1}(a) and also reported in ref.~\cite{Nehla_prb_19}. In order to calculate the order parameter for the degree of B2 ordering, S$_{\rm B2}$ can be defined using the intensity ratio of (002) and (004) peaks according to the following equation~\cite{Okamura_mt_06, Takamrua_jap_09,Wen_pra_14}: 
$S_{\rm B2} = \sqrt{[I_{002}/I_{004}]_{\rm expt}/[I_{002}/I_{004}]_{\rm calc}}$, where [$I_{002}/I_{004}$]$_{\rm expt}$ and [$I_{002}/I_{004}$]$_{\rm calc}$ are the ratio of the integrated intensity of the (002) to the (004) peaks, as determined from the experimentally observed and simulated curves, respectively. Interestingly, the order parameter S$_{\rm B2}$ is found to be $\approx$71\%, which demonstrates the high B2-type ordering in the thin film sample. From the XRD pattern, we calculate the crystallite size $D$ using the Scherrer formula~\cite{Patterson_pr_39}, D = [ $k$$\lambda$/$\beta$$\cos$$\theta$], where $k$ is a dimensionless shape factor and value is taken approximately 0.89, $\beta$ is the FWHM of the diffraction peak, $\lambda$ is the x-ray wavelength of 1.5406~\AA, and $\theta$ is the diffraction angle (in units of degree) of the (004) reflection peak, see Fig.~\ref{fig1}(b). The FWHM and crystallite size are estimated to be 0.72$\pm$0.03$^{\rm o}$ and about 13~nm, respectively.

\begin{figure}[ht]
		\includegraphics[width=3.4in]{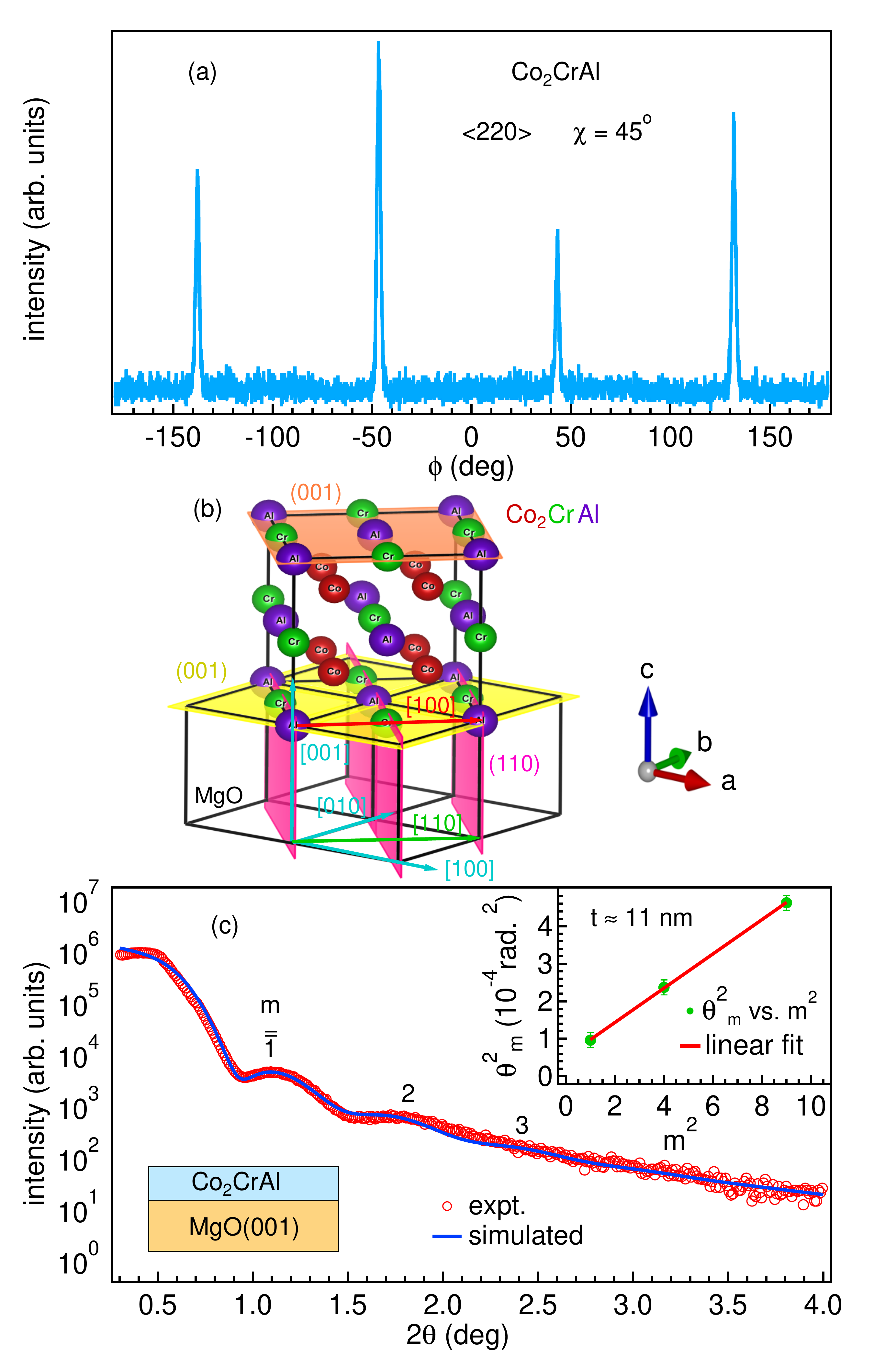}
				\caption{(a) The $\phi$ scan along the plane (220) at tilting angle $\chi$ = 45$^{\rm o}$ of Co$_2$CrAl film. (b) The schematic of the orientation of Co$_2$CrAl cubic crystal along 45$^{\rm o}$ to the edge of the MgO cubic substrate. (c) The x-ray reflectivity experimental (red circle) and fitted (blue line) curves where the inset shows the linear fitting of square of the Bragg angle ($\theta$) as a function of m$^2$, corresponding to the m$^{th}$ oscillation maxima. }
\label{fig2}
\end{figure}

It is important to understand the growth nature of the film and calculate the film thickness. If we simply consider the lattice misfit between the Co$_2$CrAl ($a=$ 5.730~\AA) bulk target and the single crystalline MgO substrate unit cell parameter ($a_{\rm sub} =$ 4.211~\AA), it gives large strain (36\%) in the film and does not favor the epitaxial growth. However, the lattice mismatch is found to be only -3.8\% when we consider the face diagonal length of MgO($\sqrt{2}$$a$$_{\rm sub}$), which is considered suitable for the epitaxial growth \cite{Kelekar_jap_04,Matsuda_jcg_06,Wang_apl_08}. Generally, the epitaxial nature of the thin films on a single crystalline substrate was established with the $\phi$ scan measurement where the sample and substrate both have the four-fold symmetry along the same plane in the equal interval of angle~\cite{Yamada_prb_19, Pan_pra_17, Kushwaha_apl_17, Kelekar_jap_04}. In order to check this, we perform the $\phi$-scan along the (220) (2$\theta$ = 45.3$^{\rm o}$ and $\chi$ = 45$^{\rm o}$) plane orientation of Co$_2$CrAl film, as shown in Fig.~\ref{fig2}(a), which confirm the four well-defined peaks (220), (202), (2$\bar{2}$0), and (20$\bar{2}$) and are periodically separated from one another with an angular difference of 90$^{\rm o}$, i.e., four-fold symmetry. These results indicate the epitaxial growth of the film on MgO (001) substrate. Thus, the relation of the epitaxial growth is defined as Co$_2$CrAl(001)[100]$\vert$$\vert$MgO(001)[110]. The visualization of growth of Co$_2$CrAl film on MgO: horizontally Co$_2$CrAl(001) planes are parallel to the MgO(001) planes and Co$_2$CrAl[100] direction is parallel to MgO[110] direction and vertically Co$_2$CrAl[100] direction is perpendicular to the MgO (110) planes. Fig.~\ref{fig2}(b) presents the schematic of orientation of Co$_2$CrAl cubic crystal along with 45$^{\rm o}$ to the edge of the MgO cubic crystal structure, where one unit cell of Co$_2$CrAl sits diagonally on the four unit cells of MgO substrate. Further, we estimate the thickness, density, and roughness of the layer of the thin film using the specularly x-rays reflectivity (XRR) data. Fig.~\ref{fig2}(c) shows the experimental data (red circle) with modelling of a single layer where the Kiessig fringes arise due to the interface formation between the Co$_2$CrAl film and the MgO substrate. A few number of fringes and a large density difference between the film and substrate for small-angle range suggest the mixed interface and higher amplitude of oscillations, respectively. Also, faster decay rate in the intensity of the XRR owing to the large surface or interface roughness \cite{Yasaka_rj_10}. We have fitted the XRR experimental data using a combined fitting algorithm that defines as a combination of the segmented and genetic algorithm (blue line) in the X’pert reflectivity software. The inset of Fig.~\ref{fig2}(c) shows the schematic model of single layer Co$_2$CrAl on MgO substrate. The obtained value of thickness, roughness, and density are inferred from the fitting as 12$\pm$1~nm, 3.85$\pm$0.15~nm, and 5.70$\pm$0.37~g/cm$^3$, respectively. Also, the thickness of the thin film is calculated from the XRR data using the modified Bragg's equation $\theta_{m}^{2}=\theta_{c}^{2}+\left(\frac{\lambda}{2 t}\right) m^{2}$~\cite{Kushwaha_apl_17, Birkholz_wiley_06} where $m$ is an integer indicates oscillation maxima, $\lambda$ is the wavelength of the x-ray Cu-K$_{\alpha}$, $t$ is the thickness of the thin film, $\theta_c$ is the critical angle (in radians) for total reflection, and $\theta_m$ is the Bragg angle (in radians) of the m$^{th}$ oscillation maxima. The thickness of the film ($\approx$11 nm) is obtained from the calculation with the above equation, which is well matched with the fitted value of XRR spectra using X’pert reflectivity software.  
	
	\begin{figure}[ht]
		\includegraphics[width=3.4in]{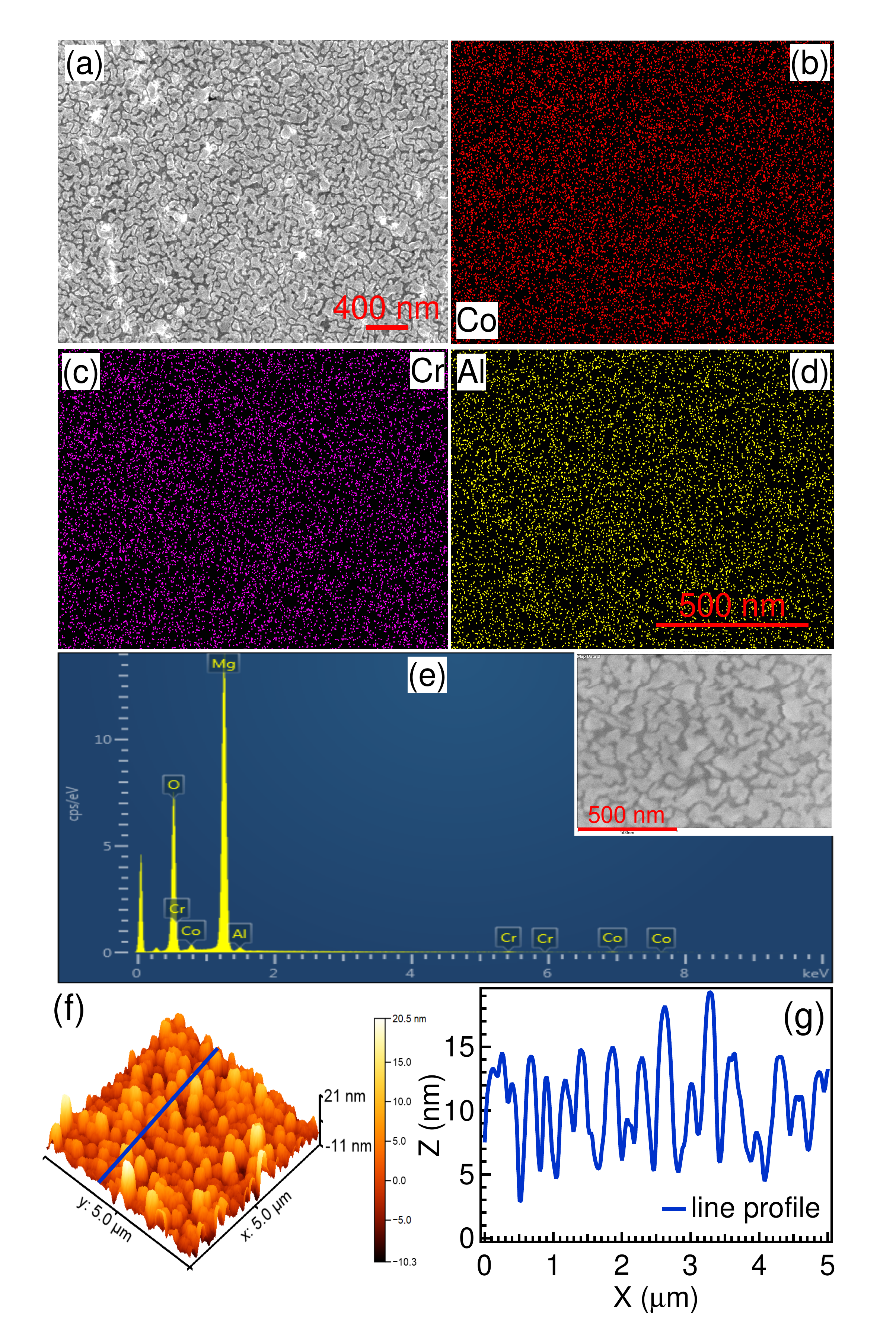}
		\caption{(a) The surface morphology is recorded using field emission secondary electron microscopy. The mapping of the elements (b) Co, (c) Cr and (d) Al show the distribution with different color coding. (e) The energy dispersive x-ray for the composition of the elements in Co$_2$CrAl film on MgO(001) substrate. (f) The surface topography image in 3-dimensional space using atomic force microscopy (AFM); solid blue line drawn over the surface, and (g) the corresponding scan profile.}
		\label{fig3}
	\end{figure}
	
In Fig.~\ref{fig3}(a), we present the FE-SEM image of Co$_2$CrAl film sample to show the surface morphology, which indicates the large fractal grains of a maze-like shape. A similar kinds of fractals are reported for CoPt film on STO(001) where the islands of arbitrary shape were found to change to nearly circular grains with increasing the growth temperature \cite{Rakshit_apl_06}. Thus the higher growth temperature is favorable for decreasing the dislocation density and improving the surface smoothness of the film. The elemental mapping images show the distribution of different elements in Co$_2$CrAl film, as shown with the red, purple, and yellow colors for Co, Cr, and Al, respectively in Figs.~\ref{fig3}(b--d). The energy dispersive x-ray (EDX) study confirms the nearly stoichiometry ratio, as shown in Fig.~\ref{fig3}(e). We found the elemental composition of  2Co:Cr:Al$\approx$56:25:20, which is nearly close to the targeted composition. The atomic force microscopy (AFM) image in Figs.~\ref{fig3}(f) presents the surface topography of the film in the scans area of 5~$\mu$m x 5~$\mu$m at room temperature in a 3-dimensional space image. A solid blue line is drawn over the surface of the AFM image to see the height profile of the islands and the corresponding scan profile. In Fig.~\ref{fig3}(g), a clear view of 3-dimensional growth with visualization of possibly islands of triangular shape are found. The obtained root mean square roughness from the AFM ($\rho$$_{rms}$) is $\approx$4~nm, which is consistent with the one obtained from XRR analysis. 

	\begin{figure}[ht]
		\includegraphics[width=3.4in]{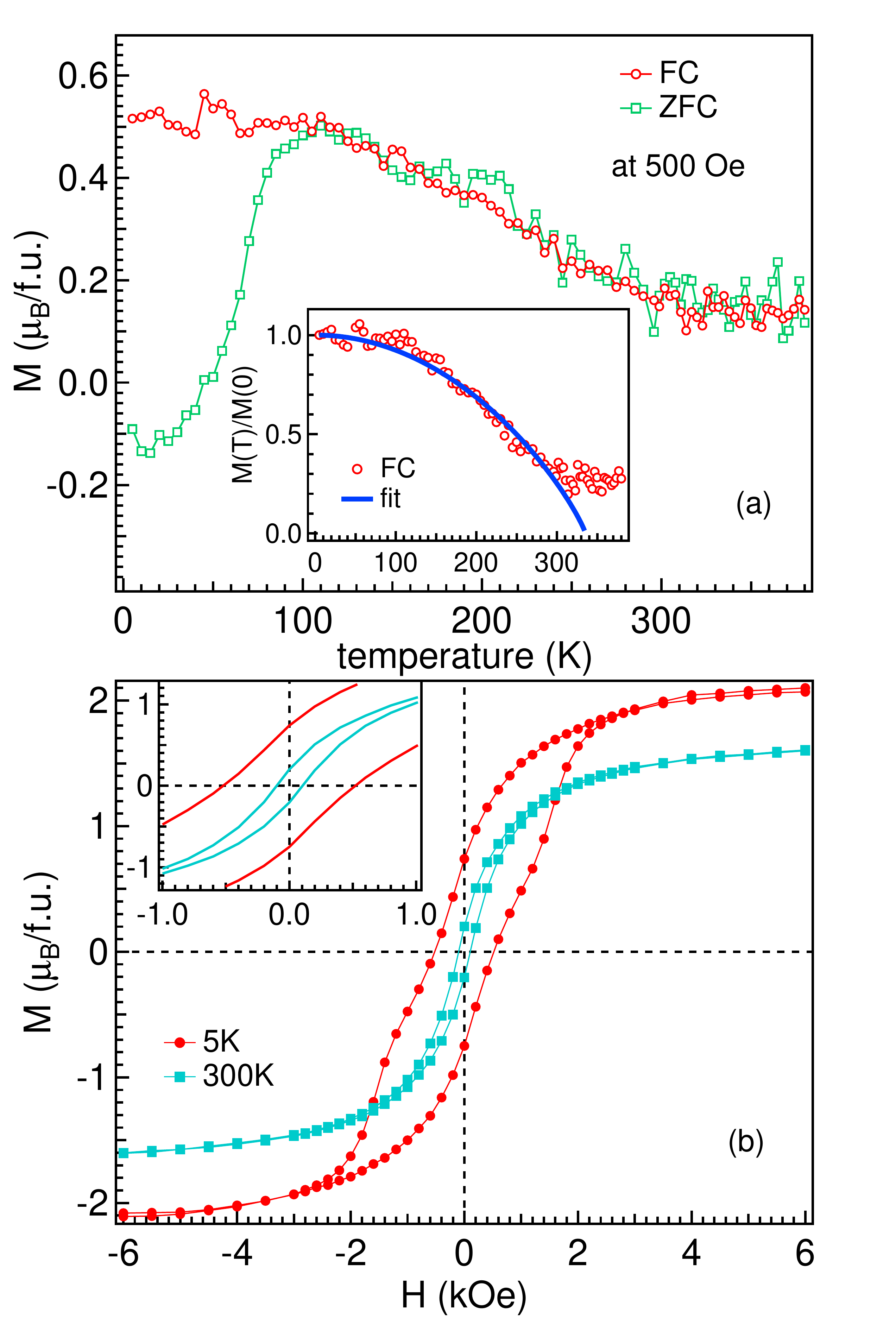}
		\caption{(a) The thermo-magnetization curves of the thin film sample, measured in the zero-field-cooled and field-cooled processes and at 500~Oe magnetic field in the temperature range of 5--300~K. The inset shows the fitting of FC data with an empirical power law (solid blue line). (b) In-plane isothermal magnetization at 5~K and 300~K of the thin film. The inset presents the enlarged view of the hysteresis of M--H in the Co$_2$CrAl thin film. All the magnetization data are shown after subtraction of the contribution from the MgO(001) substrate.}
		\label{fig4}
	\end{figure}
	
	In order to study the temperature dependence of magnetization $M$($T$), we perform the measurements in zero-field-cooled (ZFC) and field-cooled (FC) modes over the temperature range of 5--380~K at the applied external magnetic field of 500~Oe parallel to the film direction [110], as shown in Fig.~\ref{fig4}(a). In the same protocol of the magnetization, we have recorded the thermo-magnetization data of MgO substrate independently, and subtracted off from the magnetization data of the thin film. The ZFC--FC curves clearly show a bifurcation around 100~K, similar to the bulk Co$_2$CrAl sample likely due to the presence of local anti-site disorder structure, which might produce the local anti-ferromagnetism leads to decrease in the magnetization of ZFC~\cite{Nehla_prb_19, Kudryavtesev_prb_08}. The ferromagnetic transition temperature (T$_{\rm C}$) of the film can be obtained from the fitting of magnetization FC curve using a more familiar empirical power law [$\rm M(T)=M(0)\left(1-(T/T_{\mathrm{C}})^2\right )^{0.5}$] \cite{Anupam_jpdap_10, Husmann_prb_06, Feng_aip_15}, as sown in the inset of Fig.~\ref{fig4}(a). The extracted value of T$_{\rm C}=$ 335$\pm$11~K is found to be in good agreement with the reported in refs.~\cite{Kelekar_jap_04, Nehla_prb_19}. It is consistent with various half-metallic Heusler alloys where the T$^2$ dependency of magnetization is observed and can be related to Stoner excitation with itinerant-like ferromagnetism~\cite{Husmann_prb_06, Anupam_jpdap_10, Feng_aip_15}. It is noticed that the ZFC curve exhibits a small negative magnetization below $\approx$45~K temperature. Although the negative magnetization was reported for the similar Co$_2$CrAl sample in bulk and thin film, and found an increase in the negative value of magnetization as the magnetic field strength decreases. The probable reason for negative magnetization can be the magnetic inhomogeneity in the sample, or it could be due to the residual field of the superconducting magnet trapped during cooling. However, the most reliable explanation of the negative magnetization in the ZFC mode is that the magnetization pinned during the cooling, and to overcome this value it requires the sufficient thermal energy to align the magnetic moments in the field direction, which is $\approx$45~K in the present sample. 
	
	Moreover, to understand the explicit magnetic nature of the thin film, in-plane isothermal magnetization (M--H) curves of Co$_2$CrAl thin film are recorded as a function of magnetic field at 5~K and 300~K temperatures. These hysteresis curves are plotted in the magnetic field range of $\pm$6~kOe for more clarity, as shown in Fig.~\ref{fig4}(b). The coercive field increases significantly at 5~K ($\approx$522~Oe) as compared to room temperature ($\approx$102~Oe), as clearly visible in the inset of Fig.~\ref{fig4}(b). The M--H curves depict the soft ferromagnetic behavior of the thin film along the in-plane direction \cite{Gabor_prb_11}. The value of saturation magnetization (M$_{\rm S}$) of Co$_2$CrAl thin film is found to be 2.1~$\mu$$_{\rm B}$/f.u.~and 1.63~$\mu$$_{\rm B}$/f.u.~at 5~K and 300~K temperatures, respectively, which are lower than the theoretical value according to the Slater-Pauling rule (3~$\mu$$_{\rm B}$/f.u for bulk Co$_2$CrAl). It has been reported that the magnetic moment value is mainly confined at the Co (0.8~$\mu$$_{\rm B}$) and Cr (1.6~$\mu$$_{\rm B}$) atoms \cite{Kudryavtesev_prb_08, Miura_prb_04}. The reduction in the magnetization significantly depend on the degree of chemical disorder in the system \cite{Orgassa_jap_2000} and in our sample, B2-type disorder is present, which is inferred from the XRD result. Also, the lower value of magnetization saturation 1.45~$\mu$$_{\rm B}$/f.u. than the calculated has been reported in the bulk alloy \cite{Nehla_prb_19}. However, the magnetic moment values here (406.1~emu/cc at 5~K and 315.4~emu/cc at 300~K ) are well matched with the reported value of Co$_2$CrAl thin film deposited on glass and MgO substrates \cite{Kudryavtesev_prb_08, Kelekar_jap_04}. We estimated the same thickness from x-ray reflectivity and the usage of theoretical density ($\rho$ = 6.81 g/cm$^3$) for calculating the mass of the thin film may also reflect the observed difference in the saturation magnetization with respect to the values calculated in refs.~\cite{Kushwaha_apl_17, Anupam_jpdap_10}. Note that the the high magnetic moment value in thin film sample as compared to the bulk alloy, owing to the effect of high annealing temperature (700$^{\rm o}$C). Also, as seen from the composition of Co$_2$CrAl thin film where the atomic percentage of Co atoms is higher, it results in a higher magnetic moment than the experimentally reported value of stoichiometric bulk sample \cite{Nehla_prb_19}. 
	
	\begin{figure}[ht]
		\includegraphics[width=3.45in]{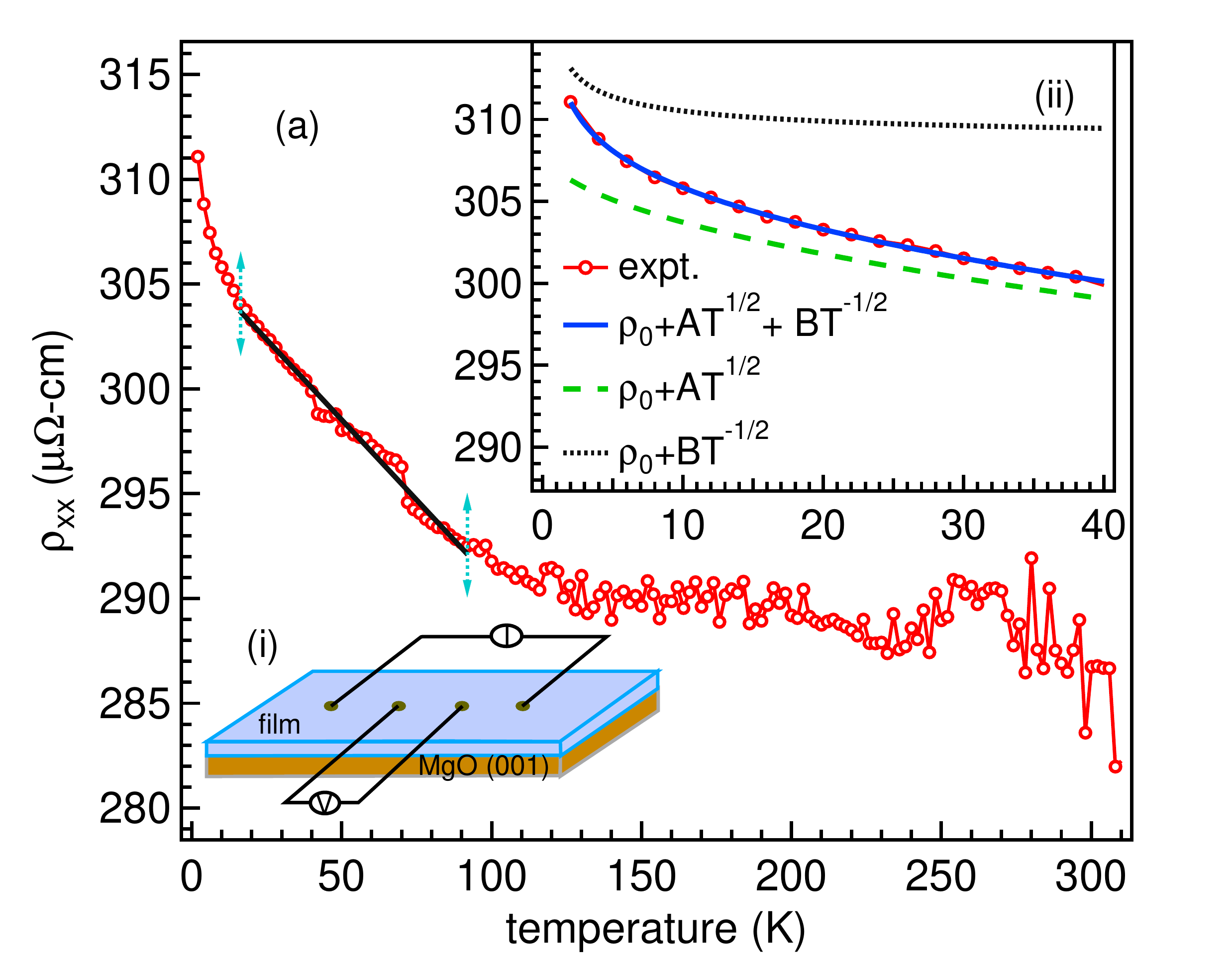}
		\caption {The {\it dc} electrical resistivity of Co$_{2}$CrAl thin film sample as a function of temperature in the range of 2--310~K during heating mode where the current is applied parallel to the film direction [110]. Inset (i) is the schematic of current and voltage connections in the four-probe method, and inset (ii) shows the low temperature resistivity data fitted with the equation ($\rho$ = $\rho$$_0$ + AT$^{1/2}$ + BT$^{-1/2}$), AT$^{1/2}$ and BT$^{-1/2}$, as shown by solid blue curve, dashed green curve and dotted black curve, respectively.}
		\label{fig5}
	\end{figure}
	
	Finally, in order to understand the transport properties, we perform the temperature dependence of dc electrical resistivity ($\rho$$_{\rm xx}$-T) in the range of 2--310~K applying 1~$\mu$A current parallel to the film direction [110] of the Co$_2$CrAl thin film grown on the MgO(001) substrate, as shown in Fig.~\ref{fig5}. The schematic of current and voltage connection in the four-probe method is shown in the inset (i) of Fig.~\ref{fig5}. The $\rho$$_{\rm xx}$ increases slowly with decrease in temperature up to around 100~K and below which we observe a significant rise in the slope, and this behavior originates due to the overlapping of the majority sub-band with Co ${3d}$ and Cr ${3d}$ at the Fermi level producing localized $3d$ electrons \cite{Kudryavtesev_prb_08}. The $\rho$$_{\rm xx}$ values are found to be $\approx$311~$\mu$$\ohm$-cm at 2~K and $\approx$285~$\mu$$\ohm$-cm at room temperature, which are relatively in agreement with the reported values of resistivity 210~$\mu$$\ohm$-cm of the bulk Co$_2$CrAl and 232~$\mu$$\ohm$-cm in the thin film grown on glass substrate \cite{Kudryavtesev_prb_08, Husmann_prb_06}. The effect of structural disorder on the resistivity values of the Co$_2$CrAl thin film as a function of temperature is reported in Refs.~\cite{Kudryavtsev_epjb_12, Kelekar_jap_04}. The  high resistivity at 300~K suggests the high B2-type ordered structure \cite{Kudryavtesev_prb_08, Kudryavtsev_epjb_12}. To extract the temperature coefficient $\alpha$, we fitted the resistivity data in 16--92~K range with a straight line, which is given by the well known general formula for TCR: $\alpha$ = (1/$\rho$)(d$\rho$/dT) \cite{Vasundhara_prb_08}. The slope (d$\rho$/dT) of the curve is found to be --0.153~$\mu$$\ohm$-cm/K, which results in the negative temperature coefficient resistivity (TCR) as --5.2$\times$10$^{-4}$ per K in the range 16--92~K, which is matched with the reported value for similar Heusler alloys \cite{Kushwaha_apl_17,Vasundhara_prb_08, Kudryavtsev_epjb_12}, which satisfies the Mooij criterion \cite{Mooij_pss_73}. This suggests that the negative TCR in the low temperature may arise either scattering of conduction electrons due to atomic disorder or localization of charge-carrier near the Fermi level \cite{Kudryavtesev_prb_08, Kelekar_jap_04}. There are several possible reasons for the origin of negative TCR reported in refs.~\cite{Kudryavtesev_prb_08, Svyazhina_jetp_13, Kudryavtsev_epjb_12}. For example, in the disorder system, the effect of the weak localization and quantum interference of electron-electron interaction gives rise to the high resistivity and negative TCR \cite{Gnida_prb_21}. In contrast, Zhang {\it et al.} reported a positive linear coefficient and relatively low value of residual resistivity \cite{Zhang_jmmm_04}. On the other hand, the negative TCR may come into effect due to the variable-range-hopping (VRH) conductivity, the Kondo effect, and Debye-Waller factor \cite{Kudryavtsev_epjb_12, Vasundhara_prb_08, Gnida_prb_21}. In the inset (ii) of Fig.~\ref{fig5}, the presence of the paramagnetic impurities in the disorder system show the sharp upturn in the resistivity data below 16~K and could not be explained with the general exponential function. Therefore, we follow the theory proposed by Altshuler and Aronov (AA) for disordered metals considering the electron-electron interactions \cite{Altshuler_elsevier_85, Gnida_prb_21}. However, the low temperature data does not follow T$^{1/2}$ relation of AA. So, an additional term is added to the equation, which can be written as $\rho$ = $\rho$$_0$ + $A$T$^{1/2}$ + $B$T$^{-1/2}$, where $A$ and $B$ are responsible for electron-electron Coulomb interaction and exchange interaction between conduction electrons and spin, respectively \cite{Gnida_prb_21, Altshuler_elsevier_85}. Although the additional correction reduces to electron-electron interaction only due to the effect of magnetic field as suggested in ref.~\cite{Altshuler_elsevier_85}. The values of $A=$ --1.5$\pm$0.05~$\mu$$\ohm$--cm K$^{-1/2}$, $B=$ 6.7$\pm$0.35~$\mu$$\ohm$--cm K$^{1/2}$ and $\rho$$_0$ = 308.4$\pm$0.2~$\mu$$\ohm$--cm are obtained from the fitting of experimental data. These values are comparable to the similar inter-metallic compounds reported in ref.~\cite{Gnida_prb_21}.
		
	\section{\noindent ~Conclusions}
	
	In summary, we have successfully grown epitaxial thin film of ferromagnetic Co$_2$CrAl Heusler alloy on MgO(001) single crystal substrate using the pulsed laser deposition. The thickness of around 12~nm was found from the analysis of x-ray reflectivity. From the analysis of field-cooled thermo-magnetization curve we extract the value of Curie temperature 335$\pm$11~K. The in-plane isothermal magnetization data exhibit the soft ferromagnetic nature where the magnetic moment is found to be 2.1~$\mu$$_{\rm B}$/f.u.~at 5~K. The temperature-dependent resistivity shows the semiconducting-like behavior with a high value of resistivity ($\approx$311~$\mu$$\ohm$-cm) suggesting the high order structure of the Co$_2$CrAl film. The growth of epitaxial and highly crystalline Heusler alloy thin films having large saturation magnetization values are considered very suitable for spintronics applications.
	
	\section{\noindent ~ACKNOWLEDGMENTS}
	GDG acknowledges MHRD, India, for fellowship through IIT Delhi. GDG also thanks Ajay Kumar, Rishabh Shukla and Dr. Priyanka Nehla for their help during the thin film deposition and useful discussion. We thank the department of physics and CRF for providing the  research facilities for measurements: XRD, PPMS EVERCOOL-II, SQUID, EDX, and AFM/MFM. RSD gratefully acknowledges the financial support from BRNS through DAE Young Scientist Research Award with project sanction No. 34/20/12/2015/BRNS. The pulsed laser deposition is financially supported by IIT Delhi through seed grant with reference no. BPHY2368 and SERB-DST through early career research (ECR) award with project reference no. ECR/2015/000159.

\end{document}